\documentclass[a4paper]{article}
\usepackage{RR}
\usepackage{hyperref,amssymb,amsmath,array}
\usepackage{algorithm,algorithmic}
\RRdate{November 2006}
\RRauthor{
Mar\'ia Naya Plasencia
  
}
\authorhead{Naya Plasencia}
\RRtitle{Cryptanalyse de Achterbahn-128/80}
\RRetitle{Cryptanalysis of Achterbahn-128/80}
\titlehead{Cryptanalysis of Achterbahn-128/80}
\RRresume{
Ce papier pr\'esente deux attaques sur Achterbahn-128/80, la derni\`ere version
d'un des algorithmes propos\'es dans  le cadre de eSTREAM. L'attaque sur la
version de 80 bits, Achterbahn-80, est en $2^{56.32}$.  L'attaque sur
Achterbahn-128 a besoin  de $2^{75.4}$ calculs et $2^{61}$  bits de suite
chiffrante. Ces  attaques sont bas\'ees sur une  am\'elioration de l'attaque
propos\'ee par Hell et Johansson  sur la version~2 d'Achterbahn et aussi sur
un algorithme qui tire profit des petites longueurs des registres.} 

\RRabstract{
This paper presents two attacks against Achterbahn-128/80, the last version
of one of the stream cipher proposals in the eSTREAM project. The attack against the
80-bit  variant,  Achterbahn-80,  has complexity  $2^{56.32}$.  The  attack
against Achterbahn-128 requires $2^{75.4}$ operations and $2^{61}$ keystream
bits.  These attacks  are based on an improvement of the  attack due to Hell
and Johansson against Achterbahn version~2 and also on an algorithm that makes profit of the short
lengths of the constituent registers.}
\RRmotcle{eSTREAM,   chiffrement   \`a   flot,  Achterbahn,   attaques   par
  corr\'elation, relations de parit\'e, cryptanalyse}
\RRkeyword{eSTREAM, stream ciphers, Achterbahn, correlation attacks, parity checks, cryptanalysis}
\RRprojet{CODES}  
\RRtheme{\THSym} 
\URRocq 

\newtheorem{definition}{Définition}

\newtheorem{lemme}[definition]{Lemma}

\newcommand{\rmPr}{\textrm{Pr}}

{\samepage \hfill$\diamond$\end{trivlist}}
{\end{trivlist}}

\begin{document}
\makeRR   

\section{Introduction}

Achterbahn~\cite{Gammel_Gottfert_Kniffler_1,Gammel_Gottfert_Kniffler_2} is a stream cipher proposal submitted to the eSTREAM project.  After the
cryptanalysis             of            the             first            two
versions~\cite{Johansson_Meier_Muller,Hell_Johansson}, it has moved on to a new one called
  Achterbahn-128/80~\cite{Gammel_Gottfert_Kniffler_3}   published   in  June
  2006. Achterbahn-128/80 corresponds to two keystream generators with key sizes of 128 bits and 80
  bits, respectively. Their maximal keystream length is limited to $2^{63}$.

We present here two attacks  against both generators. The attack against the
80  bit  variant,  Achterbahn-80,  has complexity  $2^{56.32}$.  The  attack
against Achterbahn-128 requires $2^{75.4}$ operations and $2^{61}$ keystream
bits. These attacks are based on an
improvement  of the  attack  against  Achterbahn version~2  and  also on  an
algorithm that makes profit of the short lengths of the constituent registers. 

The  paper   is  organized   as  follows.  Section   2  presents   the  main
specifications of  Achterbahn-128/80. Section  3 then describes  the general
principle of the attack proposed by Hell and Johansson~\cite{Hell_Johansson}
against the previous version of  the cipher Achterbahn v2, since our attacks
rely on a similar technique. We also exhibit a new attack against Achterbahn
v2 with  complexity $2^{49.8}$, while  the best previously known  attack had
complexity   $2^{59}$.  Section   4  then   presents  two   attacks  against
Achterbahn-80 and Achterbahn-128 respectively.

\subsection{Main specifications of Achterbahn-128}

Achterbahn-128 is a keystream generator,  consisting of 13 binary nonlinear
feedback shift registers (NLFSRs). The length of register $i$ is $L_i=21+i$ for $i=0,1,\ldots,12$. These NLFSRs are primitive
in the sense that their periods $T_i$ are equal to $2^{L_i}-1$. The sequence
which is used as an
input to the Boolean combining function  is not the output sequence of the
NLFSR directly, but a shifted version  of itself. The shift amount depends
on the register number, but it is fixed for each register. 
In the following, $x_i=\left(x_i(t)\right)_{t\geq0}$ for $0\leq i\leq 12$ denotes the shifted
version of the output of the register $i$ at time $t$.

The output of the keystream generator at time $t$, denoted by $S(t)$, is the one of the Boolean combining
function $F$ with the  inputs corresponding to the output
sequences        of        the        NLFSRs       correctly        shifted,
i.e. $S(t)=F(x_0(t),\ldots,x_{12}(t))$.  The Boolean combining function
$F$ is given by:\\

$F(x_0, x_1, . . . , x_{12}) = x_0 + x_1 + x_2 + x_3 + x_4 + x_5 + x_7 + x_9 + x_{11} + x_{12} + x_0x_5
+ x_2x_{10} + x_2x_{11} + x_4x_8 + x_4x_{12} + x_5x_6 + x_6x_8 + x_6x_{10} + x_6x_{11}
+ x_6x_{12} + x_7x_8 + x_7x_{12} + x_8x_9 + x_8x_{10} + x_9x_{10} + x_9x_{11} + x_{9}x_{12}
+ x_{10}x_{12} + x_0x_5x_8 + x_0x_5x_{10} + x_0x_5x_{11} + x_0x_5x_{12} + x_1x_2x_8
+ x_1x_2x_{12}  + x_1x_4x_{10} +  x_1x_4x_{11} + x_1x_8x_9 +  x_1x_9x_{10} +
x_1x_9x_{11} + x_1x_9x_{12} + x_2x_3x_8 + x_2x_3x_{12} + x_2x_4x_8 + x_2x_4x_{10} + x_2x_4x_{11}
+ x_2x_4x_{12} + x_2x_7x_8 + x_2x_7x_{12} + x_2x_8x_{10} + x_2x_8x_{11} + x_2x_9x_{10}
+ x_2x_9x_{11} + x_2x_{10}x_{12} + x_2x_{11}x_{12} + x_3x_4x_8 + x_3x_4x_{12} + x_3x_8x_9
+ x_3x_9x_{12} + x_4x_7x_8 + x_4x_7x_{12} + x_4x_8x_9 + x_4x_9x_{12} + x_5x_6x_8
+ x_5x_6x_{10} + x_5x_6x_{11} + x_5x_6x_{12} + x_6x_8x_{10} + x_6x_8x_{11} + x_6x_{10}x_{12}
+ x_6x_{11}x_{12} + x_7x_8x_9 + x_7x_9x_{12} + x_8x_9x_{10} + x_8x_9x_{11} + x_9x_{10}x_{12}
+ x_9x_{11}x_{12} + x_0x_5x_8x_{10} + x_0x_5x_8x_{11} + x_0x_5x_{10}x_{12} + x_0x_5x_{11}x_{12}
+ x_1x_2x_3x_8 + x_1x_2x_3x_{12} + x_1x_2x_7x_8 + x_1x_2x_7x_{12} + x_1x_3x_5x_8
+ x_1x_3x_5x_{12} + x_1x_3x_8x_9 + x_1x_3x_9x_{12} + x_1x_4x_8x_{10} + x_1x_4x_8x_{11}
+ x_1x_4x_{10}x_{12} + x_1x_4x_{11}x_{12} + x_1x_5x_7x_8 + x_1x_5x_7x_{12} + x_1x_7x_8x_9
+ x_1x_7x_9x_{12} + x_1x_8x_9x_{10} + x_1x_8x_9x_{11} + x_1x_9x_{10}x_{12} + x_1x_9x_{11}x_{12}
+ x_2x_3x_4x_8 + x_2x_3x_4x_{12} + x_2x_3x_5x_8 + x_2x_3x_5x_{12} + x_2x_4x_7x_8
+ x_2x_4x_7x_{12} + x_2x_4x_8x_{10} + x_2x_4x_8x_{11} + x_2x_4x_{10}x_{12} + x_2x_4x_{11}x_{12}
+ x_2x_5x_7x_8 + x_2x_5x_7x_{12} + x_2x_8x_9x_{10} + x_2x_8x_9x_{11} + x_2x_9x_{10}x_{12}
+ x_2x_9x_{11}x_{12} + x_3x_4x_8x_9 + x_3x_4x_9x_{12} + x_4x_7x_8x_9 + x_4x_7x_9x_{12}
+    x_5x_6x_8x_{10}    +    x_5x_6x_8x_{11}    +    x_5x_6x_{10}x_{12}    +
x_5x_6x_{11}x_{12}.$\\

Its main cryptographic properties are :

\begin{itemize}
\item balancedness
\item algebraic degree = 4
\item correlation immunity order = 8
\item nonlinearity = 3584
\item algebraic immunity = 4
\end{itemize}
 
\subsection{Main specifications of Achterbahn-80}

Achterbahn-80 consists of 11 registers, which
are the same ones as in the above case, except for the first and the last
ones. The Boolean combining function, $G$, is a sub-function of $F$ :
\begin{eqnarray*}
G(x_1,\ldots,x_{11})=F(0,x_1,\ldots,x_{11},0).
\end{eqnarray*}
Its main cryptographic properties are :

\begin{itemize}
\item balancedness
\item algebraic degree = 4
\item correlation immunity order = 6
\item nonlinearity = 896
\item algebraic immunity = 4
\end{itemize}
 
As we can see, Achterbahn-128 contains Achterbahn-80 as a substructure. 

\subsection{The key-loading algorithm}

The key-loading algorithm uses the key $K$ and an initial value $IV$. 
The method for initializing the registers is the following one:
first of all, all registers are filled with the bits of $K||IV$. After that,
register $i$ is clocked $a-L_i$ times where $a$ is the number of bits of
$K||IV$,  and  the remaining  bits  of $K||IV$  are  added  to the  feedback
bit. Then, each register outputs one bit. Those bits are taken as input
on the Boolean combining function, which  outputs a new bit. This bit is now
added to the feedbacks for $32$ additional clockings. 
Then we overwrite the last cell of  each register with a 1, in order to
avoid the all zero state. 
   
This algorithm has  been modified in relation to  the previous versions. The
aim of this modification is to  prevent the attacker from recovering the key
$K$ from the knowledge of the initial
states of some registers.

\section{Attack against Achterbahn version~2 with complexity of $\boldsymbol{2^{49.8}}$}

\subsection{Principle of Hell and Johansson attack against Achterbahn~v2}

Achterbahn version  2 was the previous  version of Achterbahn.  The main and
most important  differences to this last  one, which are used  by the attack
are that:

\begin{itemize}
\item it had 10 registers, with lengths between 19 and 32 bits,
\item the Boolean function, $f$, had correlation immunity order 5. 
\end{itemize}

This      version     has     been      broken     by      Johansson     and
Hell~\cite{Hell_Johansson}. Their attack is
a distinguishing attack that relies on the following well-known lemma, which is a
particular case of~\cite[Th. 6]{Baigneres_Junod_Vaudenay04}. 
\begin{lemme}\label{lemme1}

Let $X$ be a random variable that takes its values into ${\bf F}_2$ with a
distribution $D$ close to the uniform distribution that is
$${\rmPr}_{D}[X=1]=\frac{1}{2}(1+\varepsilon) \mbox{with}\;|\varepsilon|\ll1. $$
Then, for a number of samples
$$N=\frac{d}{\varepsilon^2}$$  where  $d$  is   a  real  number,  the  error
probability of the optimal distinguisher is approximately
$\Phi(-\sqrt{d}/2)$,  where  $\Phi$  is  the  distribution  function  of  the
standard normal distribution:
$$\Phi(x)=\frac{1}{2\pi}\int_{-\infty}^x\exp\left(-\frac{t^2}{2}\right)dt.$$

\end{lemme}

In  the following,  we will  consider $d=1$  which corresponds  to  an error
probability of about 0.3.  The previous quantity $\varepsilon$ that measures
the distance between $D$ and the uniform distribution is called the {\em bias} of $D$. 

The attack proposed by Hell and Johansson exploits a quadratic approximation $q$
of the combining function $f$: 
$$q(y_1,\ldots,y_n)= \sum _{j=1}^sy_{i_j}+\sum_{i=1}^{m}(y_{j_i}y_{k_i})$$ 
with $m$ quadratic terms and which satisfies
$$\rmPr[f(y_1,\ldots,y_n)=q(y_1,\ldots,y_n)]= \frac {1}{2}(1+ \varepsilon ).$$

We   build   the   parity-check   equations,  as   the   ones   introduced
by~\cite{Johansson_Meier_Muller}, that make disappear the quadratic terms by
summing up:
$$
q(t)= \sum_{j=1}^sx_{i_j}(t)+\sum_{i=1}^mx_{j_i}(t)x_{k_i}(t)
$$
at $2^m$ different moments $(t+ \tau)  $ moments, where $\tau$ varies in the
set of the linear combinations with $0-1$ coefficients of $T_{j_1}T_{k_1}, T_{j_2}T_{k_2}, \ldots, T_{j_m}T_{k_m}$.  In the
following,  this set  is denoted  by  $\langle T_{j_1}T_{k_1},\ldots, T_{j_m}T_{k_m}\rangle$,
i.e:
$$\langle                                              T_{j_1}T_{k_1},\ldots,
 T_{j_m}T_{k_m}\rangle=\left\{\sum_{i=1}^nc_iT_{j_ik_i}, c_1,\ldots,c_m \in\{0,1\}\right\}.$$
 This leads to
\begin{eqnarray*}
pc(t) & = & \sum_{\tau \in \langle T_{j_1}T_{k_1}, \ldots, T_{j_m}T_{k_m} \rangle}q(t+
\tau )\\
& = & \sum_{\tau \in \langle T_{j_1}T_{k_1}, \ldots, T_{j_m}T_{k_m}
\rangle} \left( x_{i_1}(t+ \tau ) + \ldots+ x_{i_s}(t+ \tau)\right). 
\end{eqnarray*} 
We then decimate the sequence $\left(pc(t)\right)_{t\geq0}$ by the periods of $p$ sequences among
$(x_{i_1}(t))_{t\geq0},\ldots,(x_{i_s}(t))_{t\geq0}$.  We  can suppose  here
without loss of generality that the  periods of the first $p$ sequences have
been chosen.

Now a new parity-check, $pc_p$, can be defined by:
$$pc_p(t) = pc(t T_{i_1}\ldots T_{i_p}) .$$ 
This way, the influence of those $p$ registers on the parity-check $pc_p(t)$
 corresponds to the addition of a constant for all $t\geq0$, so it will be $0$ or $1$ for all the parity-checks.  

Now, the attack consists in  performing an exhaustive search for the initial
states  of   the  $(s-p)$  remaining   registers,  i.e.  those   of  indices
$i_{p+1},\ldots,i_s$. For  each possible values for these  initial states, we
compute: 
\begin{eqnarray}
\label{eq1}
\sigma(t)=\sum_{\tau\in  \langle  T_{j_1}T_{k_1},  \ldots,  T_{j_m}T_{k_m}
  \rangle}\left[S(tT_{i_1}\ldots
  T_{i_p}+\tau)+\sum_{j=p+1}^sx_{i_j}(tT_{i_1}\ldots T_{i_p}+\tau)\right] 
\end{eqnarray}
We have
$$\rmPr[\sigma(t)=0]=\frac {1}{2}(1+ \varepsilon^{2^m} ).$$
Using this bias,  we can distinguish the keystream  $(S(t))_{t\geq0}$ from a
random sequence and also recover the initial states of $(s-p)$ constituent registers.

\subsection{Complexity}

\begin{itemize}
\item We will have $2^m$ terms in each parity-check. That means that we need
  to  compute  $\varepsilon^{-2^{m+1}}=2^{n_b2^{m+1}}$  values   of  $\sigma(t)$  for  mounting  the
  distinguishing attack, where $n_b= \log_2\varepsilon^{-1}$. Besides, $\sigma(t)$
  is defined by ($\ref{eq1}$), implying that the attack requires 
 $$2^{n_b2^{m+1}+\sum_{j=1}^pL_{i_j}}+\sum_{i=1}^m2^{L_{j_i}+L_{k_i}}
\mbox{ keystream bits},$$ where $L_{i_j}$ are the lengths of the registers associated
  to the periods  by which we have decimated, and  the last term corresponds
  to the maximal distance between the bits involved in each parity-check.

\item Time complexity will be
$$2^m2^{n_b  2^{m+1} +\sum_{j=p+1}^sL_{i_j}}$$ where  $i_{p+1},\ldots,i_s$ are
		  the indices of  the registers by which period  we have not
		  decimated,  so the  registers over  whom we  have  made an
		  exhaustive search and whose initial state we are going to find.
	
\end{itemize}

\subsection{Example with Achterbahn version 2}

Hell and Johansson~\cite{Hell_Johansson} have used this attack against Achterbahn
version 2 with the following quadratic approximation:
$$Q(x_1, \ldots, x_{10})=x_1+x_2+x_3x_8+x_4x_6 .$$ 
Then, they decimate by the period of the second register, whose length is
$22$. After  that, they make an  exhaustive search over  the first register,
whose length is $19$. Time complexity will be $2^{62}$ and data complexity
$2^{59.02}$.
Using the small lengths of the registers,  time complexity can be
reduced below data complexity, so the final complexity of the attack will be $2^{59.02}$.

\subsection{Improvement of the Attack against Achterbahn version 2}

We   are  going  to   improve  the   previously  described   attack  against
Achterbahn~v2 and we reduce the complexity to $2^{49.8}$.

	For this  attack, we  use the idea  of associating the  variables in
	order to reduce the number of  terms that we will have in the parity-checks. The only effect that  this could have on the final complexity
	of the attack is to enlarge the number of required keystream bits; but  being careful,  we make  it stay  the  same while
	reducing the time complexity. 

\paragraph{The chosen approximation.}
At first, we searched between  all the quadratics approximations of $f$ with
one and two quadratic terms, as
the original attack presented by Hell and Johansson was based on a quadratic approximation. 
Finally, after looking  after a trade-off between the  number of terms, the
number of variables, the
bias... we found that none quadratic approximation was better for this
attack  than linear ones.  It is  worth noticing  that, since  the combining
function $f$  is 5-resilient, any approximation  of $f$ involves  at least 6
input   variables.  Moreover,   the   highest  bias   corresponding  to   an
approximation of $f$  by a 6-variable function is achieved  by a function of
degree one as proved in~\cite{Canteaut_Trabbia_00}. 
After analyzing all linear approximations of the Boolean combining function, we
found that the best one was:
\begin{eqnarray*}
g(x_1,\ldots,x_{10})=x_8+x_6+x_4+x_3+x_2+x_1 .
\end{eqnarray*}

We have $f(x_1,\ldots,x_{10})=g(x_1,\ldots,x_{10})$ with a probability of $\frac{1}{2}(1+2^{-3}).$

	\paragraph{Parity-checks.} 

Let us build a parity-check as follows:
\begin{eqnarray*}
ggg(t)=g(t)+g(t+T_1T_8)+g(t+T_2T_6)+g(t+T_1T_8+T_2T_6),
\end{eqnarray*}
with
$$g(t)=x_8(t)+x_6(t)+x_4(t)+x_3(t)+x_2(t)+x_1(t) .$$
The terms $x_8$,  $x_6$, $x_2$, $x_1$ will disappear and,  so, $ggg(t)$ is a
sequence that depends uniquely on the sequences $x_3$ and $x_4$. 
Adding four times the approximation has the effect of multiplying the bias
four times, so the bias of
\begin{eqnarray*}
\sigma(t)=S(t)+S(t+T_1T_8)+S(t+T_2T_6)+S(t+T_1T_8+T_2T_6)
\end{eqnarray*}
is $2^{-3\times 4}=2^{-12}$ because $4$ is the number of terms in
 $ggg(t)$.  That means  that  we will  need  $2^{3\times 4\times  2}=2^{24}$
 values of the parity-check for detecting this bias. 
 If we decimate $ggg(t)$ by the period of register $3$, we will need
\begin{eqnarray*}
2^{24}T_3+T_1T_8+T_2+T_6 &=&  2^{24+23}+2^{29+19}+2^{27+22}\\
&=& 2^{49.8} \mbox{~ bits of keystream,}
\end{eqnarray*}
and time complexity will be
$$2^{24}\times 2^{L_4}=2^{49}$$ as we only guess the initial
state of register 4.

We consider that the total complexity is given by the data complexity, as it
is higher than the time  complexity. This complexity is $2^{49.8}$ while the
complexity of the previous attack was equal to $2^{59}$.

\section{Cryptanalysis of Achterbahn-128/80}
Now, we describe a new attack against Achterbahn-80 with a complexity of
$2^{56.32}$  where   a  linear  approximation  of  the   output  function  is
  considered.  The attack is  a distinguishing  attack but  it also  allows to
  recover the initial states of certain constituent 
registers.  We also describe an attack against Achterbahn-128 with a
complexity of  $2^{75.4}$ where  we consider a  linear approximation  of the
output function  and we make  profit of the  short lengths of  the registers
involved in the proposed stream cipher.

\subsection{Cryptanalysis of Achterbahn-80}

This  attack  is very  similar  to the  improvement  of  the attack  against
Achterbahn version 2 which has been described in the previous section. 

Our attack exploits the following linear approximation of the combining function $G$:
\begin{eqnarray*}
\ell(x_1, \ldots, x_{11})=x_1+x_3+x_4+x_5+x_6+x_7+x_{10}. 
\end{eqnarray*}
Since $G$ is 6-resilient, $\ell$ is the best approximation by a 7-variable function.

For    $\ell(t)=x_1(t)+x_3(t)+x_4(t)+x_5(t)+x_6(t)+x_7(t)+x_{10}(t)$,    the
keystream      $\left(S(t)\right)_{t\geq      0}$      satisfies      $\rmPr
[S(t)=\ell(t)]=\frac{1}{2}(1-2^{-3})$.

\paragraph{Parity-checks. }

Let us build a parity-check as follows:
$$\ell\ell(t)=\ell(t)+\ell(t+T_4T_7)+\ell(t+T_6T_5)+\ell(t+T_4T_7+T_6T_5).$$
 The terms containing the sequences $x_4$, $x_5$, $x_6$, $x_7$
vanish in $\ell\ell(t)$, so $\ell\ell(t)$  depends
exclusively on the sequences $x_1$, $x_3$ and $x_{10}$. 

Adding four times the approximation has the effect of multiplying the bias
four times, so the bias of
$$\sigma(t)=S(t)+S(t+T_7T_4)+S(t+T_6T_5)+S(t+T_7T_4+T_6T_5)$$
where $(S(t))_{t\geq0}$  is the keystream, is $2^{-4\times3}$. 
This means that we need $2^{3\times4\times2}=2^{24}$ parity-checks $\sigma(t)$
to detect this bias.

	We now decimate $\sigma(t)$ by the period of the register $10$,
	which is involved in the parity-check, so we create like this a new
	parity-check: 
\begin{eqnarray*}
\sigma'(t)=\sigma(t(2^{31}-1)).
\end{eqnarray*}
Then, the  attack performs  an exhaustive search  for the initial  states of
registers $1$ and $3$. Its time complexity is $2^{24}\times 2^{L_1+L_3}=2^{70}$.

The number of keystream bits that we need is $$2^{24}\times T_{10}+T_4T_7+T_6T_5=2^{56.32}$$. 

\subsection{Cryptanalysis of Achterbahn-128}

Now,  we present  a distinguishing  attack against  the 128-bit  version of
Achterbahn which also recovers the initial states of two registers. 

We consider the following approximation of the combining function $F$:
\begin{eqnarray*}
\ell(x_0, \ldots, x_{12})=x_0+x_3+x_7+x_4+x_{10}+x_8+x_{9}+x_1+x_2. 
\end{eqnarray*}
 Then,                                                                    for
 $\ell(t)=x_0(t)+x_3(t)+x_7(t)+x_4(t)+x_{10}(t)+x_8(t)+x_{9}(t)+x_1(t)+x_2(t)$,
 we have $\rmPr[S(t)=\ell(t)]=\frac {1} {2}(1+2^{-3}).$

\paragraph{Parity-checks. }

The period of any sequence obtained by combining the registers $0$, $3$ and $7$
is equal  to lcm$(T_0,T_3,T_7)$,  i.e. $2^{59.3}$ as  $T_0$ $T_3$  and $T_7$
have common divisors. We are going to denote this value by $T_{0,3,7}$. 

If we build a parity check as follows:
$$\ell\ell\ell(t)=\sum_{\tau\in\langle T_{0,3,7},T_{4,10},T_{8,9}\rangle}\ell(t+\tau)
,$$the terms  containing the  sequences $x_0$, $x_3$,  $x_7$, $x_4$, $x_{10}$,  $x_8$, $x_{9}$
will disappear from $\ell\ell\ell(t)$,  so $\ell\ell\ell(t)$ depends
exclusively on the sequences $x_1$ and $x_2$:
\begin{eqnarray*}
\ell\ell\ell(t)                    &                   =                   &
\sum_{\tau\in\langle T_{0,3,7},T_{4,10},T_{8,9}\rangle}\ell(t+\tau)\\
& = & \sum_{\tau\in\langle T_{0,3,7},T_{4,10},T_{8,9}\rangle} x_1(t+\tau)+
 x_2(t+\tau)\\
& = & \sigma_1(t) + \sigma_2(t),\\
\end{eqnarray*}
where  $\sigma_1(t)$  and  $\sigma_2(t)$  are  the parity-checks  calculated  on  the
sequences generated by NLFSRs 1 and 2. 

Adding eight times the approximation has the effect of multiplying the bias
eight times, so the bias of
$$\sigma(t)=\sum_{\tau\in\langle T_{0,3,7},T_{4,10},T_{8,9}\rangle}S(t+\tau)$$
where $(S(t))_{t\geq0}$  is the keystream, is $2^{-8\times3}$. 
So:
$$\rmPr[\sigma(t) + \sigma_1(t) + \sigma_2(t)= 1]=\frac {1} {2} (1-\varepsilon^8).$$ 
This means that we need $2^{3\times8\times2}=2^{48}$ values of $\sigma(t)+\sigma_1(t)+\sigma_2(t)$
to detect this bias.

We now describe an algorithm for computing the sum $\sigma(t) + \sigma_1(t) +
\sigma_2(t)$ over all  values of $t$. This algorithm  has a lower complexity
than an  exhaustive search for the  initial states of the  registers $1$ and
$2$ simultaneously. 
Here we use $(2^{48}-2)$ values of $t$ since $(2^{48}-2)=T_2\times(2^{25}+2).$

We can write it down as follows:
\begin{eqnarray*}
\sum_{t'=0}^{2^{48}-3}\sigma(t')\oplus\ell\ell\ell(t')  & = &
\sum_{k=0}^{T_2-1}\sum_{t=0}^{2^{25}+1}\sigma(T_2t+k)\oplus\ell\ell\ell(T_2t+k)\\
&   =   &  \sum_{k=0}^{T_2-1}\sum_{t=0}^{2^{25}+1}\sigma(T_2t+k)\oplus
  \sigma_1(T_2t+k) \oplus \sigma_2(T_2t+k)\\
&                                     =                                    &
\sum_{k=0}^{T_2-1}\left[(\sigma_2(k)\oplus1)\left(\sum_{t=0}^{2^{25}+1}\sigma(T_2t+k)   \oplus
    \sigma_1(T_2t+k)\right)\right.+\\
& &\left.\sigma_2(k)\left((2^{25}+2)-\sum_{t=0}^{2^{25}+1}\sigma(T_2t+k)\oplus \sigma_1(T_2t+k)\right)\right], \\
\end{eqnarray*}
since $\sigma_2(T_2t+k)$ is constant for a fixed value of $k$. 

At this point, we can obtain $\sigma(t)$  from the keystream and we can make an
exhaustive search for the initial state of register 1. More precisely:

\begin{itemize}
 \item We choose  an initial state for register 2, e.g.  the all one initial
 state. We compute and save a binary vector $V_2$ of length $T_2$:
$$V_2[k]=\sigma_2(k),$$
where the sequence $x_2$ is generated from the choosen initial state. The complexity of this state  is $T_2\times2^3$ operations.
\item For each possible initial state of register 1:
      \begin{itemize}
      \item we compute and save a vector $V_1$ composed of $T_2$ integers of 26 bits. 
$$V_1[k]=\sum_{t=0}^{2^{25}+1}\sigma(T_2t+k) \oplus \sigma_1(T_2t+k).$$
The complexity of this state is:
$$2^{48}\times(2^4+2^{4.7})=2^{53.4} $$ 
 for  each possible  initial state  of register  1,  where $2^4$  corresponds to  the
  number     of      operations     required     for      computing     each
  $\left(\sigma(t)+\sigma_1(t)\right)$ and $(2^{25}+2)\times2^{4.7}=(2^{25}+2)\times26$
  is the cost of summing up $2^{25}+2$ integers of 26 bits.
 
       \item For each possible $i$ from $0$ to $T_2-1$:
              \begin{itemize}
               \item we define $V_2'$ of length $T_2$:
$$V_2'[k]=V_2[k+i \mod T_2].$$
Actually,        $\left(V_2'[k]\right)_{k<T_2}$        corresponds        to
$\left(\sigma_2(k)\right)_{k<T_2}$  when  the initial  state  of register  2
corresponds to internal state after clocking $R2$ $i$ times from the all one
initial state.
                \item With the two vectors that we have obtained, we compute:
\begin{eqnarray}
\label{eq2}
\sum_{k=0}^{T_2-1}\left[\left(V_2'[k]\oplus1\right)V_1[k]                    +
  V_2'[k]\left(2^{25}+2-V_1[k] \right)\right]. 
\end{eqnarray}

\end{itemize}
\end{itemize}
\end{itemize}

When we do this with the correct initial states of registers 1 and 2, we will find the expected bias. 
\begin {table}[!h]
\begin{center}
\fbox{
\begin{minipage}{13cm}

\begin{algorithmic}
\FOR{each possible initial state of $R1$}
\FOR{$k=0$ to $T_2-1$}
\STATE $V_1[k]=\sum_{t=0}^{2^{25}+1}\sigma(T_2t+k) \oplus \sigma_1(T_2t+k)$
\ENDFOR
\FOR{each possible initial $i$ state of $R2$}
\FOR{$k=0$ to $T_2-1$}
\STATE $V_2'[k]=V_2[k+i \mod T_2]$
\ENDFOR
\STATE $\sum_{k=0}^{T_2-1}\left[\left(V_2'[k]\oplus1\right)V_1[k]+ V_2'[k]\left(2^{25}+2-V_1[k] \right)\right]$
\IF{we find the bias}
\STATE return the initial states of $R1$ and $R2$
\ENDIF
\ENDFOR
\ENDFOR
\end{algorithmic}

\end{minipage}} 
\caption{Algorithm for finding the initial states of registers 1 and 2}\end{center}
\end{table}

The total time complexity of the attack is going to be:
$$T_1\times \left[2^{48}\times\left(2^4+2^{4.7}\right)+T_2\times2\times T_2\times 2^{4.7}\right]+T_2\times2^3=2^{75.4},$$
where $2\times T_2\times2^{4.7}$ is the time it takes to compute the sum
described by~(\ref{eq2}). 
Actually, we  can speed up the process by rewriting the  sum (\ref{eq2}) in
the following way
$$
\sum_{k=0}^{T_2-1}(-1)^{V_2[k+i]} \left(V_1[k] -
\frac{2^{25}+2}{2}\right) + T_2 \frac{2^{25}+2}{2}
$$
The issue is now  to find the $i$ that maximizes this  sum, this is the same
as  computing the  maximum of  the crosscorrelation of  two  sequences of
length $T_2$. We  can do that efficiently using a  fast Fourier transform as
explained in~\cite[pages  306-312]{Blahut}. The final complexity  will be in
$O(T_2\log T_2)$. Anyway, this does not change our total complexity as the higher term is the first one. 

The complexity is going to be, finally:
$$T_1\times \left[2^{48}\times\left(2^4+2^{4.7}\right)+O(T_2\log T_2)\right]+T_2\times2^3=2^{75.4}.$$
The length of keystream needed is:
$$T_{0,3,7}+T_{4,10}+T_{8,9}+2^{48}<2^{61} \mbox{ bits.  } $$

\section{Conclusion} 
 
We have proposed  an attack against Achterbahn-80 in  $2^{70}$. To this attack
we can apply the same algorithm as the one described in Section 3.2 against Achterbahn-128, and its time complexity will be reduced to
about $2^{45}$,  so we can consider as the total complexity the  length of the
keystream needed,  since it is bigger.  The complexity of  the attack against
Achterbahn-80 will then be $2^{56.32}$. 
An attack against  Achterbahn-128 is also proposed in  $2^{75.4}$ where fewer
than $2^{61}$ bits  of keystream are required. The  complexities of the best
attacks against all  versions of Achterbahn are summarized  in the following
table: 
\begin{table}[h]
 \begin{center} 
\setlength{\extrarowheight}{2pt}
\begin{tabular}{|c|c|c|c|}

\hline 
version & data complexity & time complexity & references\\ 
\hline\hline

  v1 (80-bit) & $2^{32}$ & $2^{55}$ & \cite{Johansson_Meier_Muller}\\ 
\hline

  v2 (80-bit) & $2^{59.02}$ & $2^{62}$ & \cite{Hell_Johansson}\\ 
\hline\hline

  v2 (80-bit) & $2^{49.8}$ & $2^{49}$ &\\ 
\hline

  v80 (80-bit) & $2^{56.32}$ & $2^{46}$ & \\ 
\hline

  v128 (128-bit) & $2^{60}$ & $2^{75.4}$ & \\ 
\hline

\end{tabular}
\caption{Attacks complexities against all versions of Achterbahn}
\end{center}
\end{table}
\bibliographystyle{plain}
\bibliography{stream}

\tableofcontents

\end{document}